\documentclass[aps,prl,twocolumn,amssymb,superscriptaddress]{revtex4-2}

\usepackage{mathtools} 
\usepackage{graphicx}
\usepackage{epstopdf}
\usepackage{amsmath}
\usepackage{dcolumn}
\usepackage{bm}
\usepackage{bbold}
\usepackage{amssymb}
\usepackage{mathrsfs}
\usepackage{amsfonts}
\usepackage{braket}
\usepackage{tabularx}
\usepackage{setspace}
\usepackage[caption=false]{subfig}

\usepackage{float}
\usepackage{array}
\usepackage[dvipsnames]{xcolor}
\newcolumntype{L}[1]{>{\raggedright\let\newline\\\arraybackslash\hspace{0pt}}m{#1}}
\newcolumntype{C}[1]{>{\centering\let\newline\\\arraybackslash\hspace{0pt}}m{#1}}
\newcolumntype{R}[1]{>{\raggedleft\let\newline\\\arraybackslash\hspace{0pt}}m{#1}}

\begin{document}
\begin{flushright}
		DESY-22-185\\
\end{flushright}

\preprint{APS/123-QED}

\title{Sensitivity of Resonant Axion Haloscopes to Quantum Electromagnetodynamics}

\author{Michael E. Tobar}
\email{michael.tobar@uwa.edu.au}
\affiliation{Quantum Technologies and Dark Matter Labs, Department of Physics, University of Western Australia, 35 Stirling Highway, Crawley, WA 6009, Australia.}
\author{Catriona A. Thomson}
\affiliation{Quantum Technologies and Dark Matter Labs, Department of Physics, University of Western Australia, 35 Stirling Highway, Crawley, WA 6009, Australia.}
\author{Benjamin T. McAllister}
\affiliation{Quantum Technologies and Dark Matter Labs, Department of Physics, University of Western Australia, 35 Stirling Highway, Crawley, WA 6009, Australia.}
\author{Maxim Goryachev}
\affiliation{Quantum Technologies and Dark Matter Labs, Department of Physics, University of Western Australia, 35 Stirling Highway, Crawley, WA 6009, Australia.}
\author{Anton V. Sokolov}
\affiliation{Department of Mathematical Sciences, University of Liverpool, Liverpool, L69 7ZL, United Kingdom}
\author{Andreas Ringwald}
\affiliation{Deutsches Elektronen-Synchrotron DESY, Notkestr. 85, 22607 Hamburg, Germany}

\begin{abstract}
Recently interactions between putative axions and magnetic monopoles have been revisited by two of us \cite{SokolovMonopole22}. It has been shown that significant modifications to conventional axion electrodynamics arise due to these interactions, so that the axion-photon coupling parameter space is expanded from one parameter $g_{a\gamma\gamma}$ to three $(g_{a\gamma\gamma},g_{aAB},g_{aBB})$. We implement Poynting theorem to determine how to exhibit sensitivity to  $g_{aAB}$ and $g_{aBB}$ using resonant haloscopes, allowing new techniques to search for axions and a possible indirect way to determine if magnetically charged matter exists. 
\end{abstract}


\maketitle

\section{Introduction}

Axions are putative particles, thought to exist because they solve the strong CP problem \cite{PQ1977,Wilczek1978,Weinberg1978,wisps,K79,Kim2010,Zhitnitsky:1980tq,DFS81,SVZ80}, as well as being prime candidates for cold dark matter \cite{Dine1983,Preskill1983,Sikivie1983,Sikivie1983b,Co2020,Co2020b,Co2021,Sikivie2021}. In this work we implement Poynting theorem \cite{tobar2021abraham} to calculate the generalised sensitivity of resonant haloscopes from axion modified electrodynamics, which includes the usual two photon anomaly coupling parameter, $g_{a\gamma\gamma}$, as well as two other electromagnetic anomaly coupling parameters, $g_{aAB}$ and $g_{aBB}$ if magnetic monopoles exist, as suggested by implementing Quantum Electromagnetodynamics (QEMD) \cite{SokolovMonopole22, Cabibbo1962, Zwanziger1971}. We conceptualise resonant experiments with either background DC or AC magnetic and electric fields, and show various configurations that can put limits on the axion coupling parameters, while also defining the equivalent experimental form factors. 

\section{Axion Modified Electrodynamics from QEMD}

The generalised axion electrodynamics, expanded to include all possible couplings between axion and electromagnetic field, was shown to be given by (in SI units),
\begin{equation}
\begin{aligned}
\vec{\nabla} \cdot \vec{E}_{1}=g_{a\gamma\gamma}c\vec{B}_{0} \cdot \vec{\nabla}a-g_{aAB} \vec{E}_{0} \cdot \vec{\nabla} a+\epsilon_0^{-1}\rho_{e1}
\end{aligned}
\label{GausMP}
\end{equation}
\begin{equation}
\begin{aligned}
\frac{\vec{\nabla} \times \vec{B}_{1}}{\mu_0}&
=\epsilon_0\partial_t{\vec{E}}_{1}+\vec{J}_{e1}+g_{a\gamma\gamma}c\epsilon_0\left(-\vec{\nabla} a\times\vec{E}_{0} -\partial_t{a} \vec{B}_{0}\right)\\
&+g_{a A B}\epsilon_0\left( -\vec{\nabla} a\times c^2\vec{B}_{0}+\partial_t{a} \vec{E}_{0}\right),
\end{aligned}
\label{AmpereMP}
\end{equation}
\begin{equation}
\begin{aligned}
\vec{\nabla} \cdot \vec{B}_{1}=-\frac{g_{a B B}}{c} \vec{E}_{0} \cdot \vec{\nabla} a+g_{a A B} \vec{B}_{0} \cdot \vec{\nabla} a,
\end{aligned}
\label{MGaussMP}
\end{equation}
\begin{equation}
\begin{aligned}
\vec{\nabla} \times \vec{E}_{1}&=-\partial_t{\vec{B}}_{1} +\frac{g_{aBB}}{c}\left(c^2\vec{\nabla}a\times\vec{B}_{0}-\partial_t{a} \vec{E}_{0}\right) \\
&+g_{aAB}\left(\vec{\nabla}a\times\vec{E}_{0}  +\partial_t{a} \vec{B}_{0}\right),
\end{aligned}
\label{FaradayMP}
\end{equation}
where $g_{aBB}$ and $g_{a A B}$ are additional axion couplings, which were shown to dominate the effective axion currents over the conventional term, $g_{a\gamma\gamma}$ \cite{SokolovMonopole22}. Here, $\vec{B}_{0}$ and $\vec{E}_{0}$ are the electromagnetic background fields, which are generated from free charge and current densities defined by $\vec{J}_{e_0}$ and $\rho_{e_0}$, and to zeroth order in axion coupling satisfy the non-modified Maxwell's equations given by, 
\begin{equation}
\begin{aligned}
&\vec{\nabla}\times \vec{B}_0=\mu_0\epsilon_0\partial_{t} \vec{E}_0+\mu_0\vec{J}_{e_0}~~~~~~\vec{\nabla}\cdot \vec{B}_0=0 \\
&\vec{\nabla}\times \vec{E}_0=-\partial_{t} \vec{B}_0~~~~~\vec{\nabla}\cdot \vec{E}_0=\epsilon_0^{-1}\rho_{e_0},
\end{aligned}
\label{Bground}
\end{equation}
while $\vec{B}_{1}$ and  $\vec{E}_{1}$ are the axion generated fields of the first order in axion couplings to photons, which will also generate the associated free current and charge densities, $\vec{J}_{e_1}$ and $\rho_{e_1}$, respectively, within the haloscope detector.

For cold dark matter it is usual to assume $\vec{\nabla}a=0$, and in this limit (\ref{GausMP})-(\ref{FaradayMP}) becomes,
\begin{equation}
\begin{aligned}
\vec{\nabla} \cdot \vec{E}_{1}=\epsilon_0^{-1}\rho_{e1},
\end{aligned}
\label{GausMPa}
\end{equation}
\begin{equation}
\begin{aligned}
\frac{\vec{\nabla} \times \vec{B}_{1}}{\mu_0}&=\epsilon_0\partial_t{\vec{E}}_{1}+\vec{J}_{e1}+(g_{a A B}\epsilon_0\vec{E}_{0}-g_{a\gamma\gamma}c\epsilon_0\vec{B}_{0})\partial_t{a},
\end{aligned}
\label{AmpereMPa}
\end{equation}
\begin{equation}
\begin{aligned}
\vec{\nabla} \cdot \vec{B}_{1}=0,
\end{aligned}
\label{MGaussMPa}
\end{equation}
\begin{equation}
\begin{aligned}
\vec{\nabla} \times \vec{E}_{1}=-\partial_t{\vec{B}}_{1}+(g_{aAB} \vec{B}_{0}-\frac{g_{aBB}}{c}\vec{E}_{0})\partial_t{a}.
\end{aligned}
\label{FaradayMPa}
\end{equation}
For this set of equations we notice the axion electric displacement current is generalised to,
\begin{equation}
\begin{aligned}
\vec{J}_{ea}=(g_{a A B}\epsilon_0\vec{E}_{0}-g_{a\gamma\gamma}c\epsilon_0\vec{B}_{0})\partial_t{a},
\end{aligned}
\label{AxCurr}
\end{equation}
and an axion magnetic displacement current exists,
\begin{equation}
\begin{aligned}
\vec{J}_{ma}=(\frac{g_{aBB}}{c}\vec{E}_{0}-g_{aAB} \vec{B}_{0})\partial_t{a},
\end{aligned}
\label{AxMagCurr}
\end{equation}
Thus, in QEMD there are three extra axion current terms due to $g_{a A B}$ and $g_{aBB}$ compared to standard axion electrodynamics.

We may also use the method of writing the modifications as effective polarizations and magnetizations \cite{TobarModAx19,TOBAR2020,SokolovMonopole22,sokolov2022gravitational,domcke2022novel}, by implementing the following vector identities, $\vec{\nabla}a\cdot\vec{X}=\vec{\nabla}\cdot(a\vec{X})-a(\vec{\nabla}\cdot\vec{X})$ and $\vec{\nabla}a\times\vec{X}=\vec{\nabla}\times(a\vec{X})-a(\vec{\nabla}\times\vec{X})$, to eqns.(\ref{GausMP})-(\ref{FaradayMP}). In this case Gauss' and Ampere's laws become,
\begin{equation}
\begin{aligned}
\vec{\nabla} \cdot(\epsilon_0\vec{E}_{1}+\vec{P}_{e1})=g_{AB}a\rho_{e0}+\rho_{e1},
\end{aligned}
\label{GausMP2}
\end{equation}
\begin{equation}
\begin{aligned}
\vec{\nabla} \times(\frac{1}{\mu_0}\vec{B}_{1}-\vec{M}_{e1})-\partial_t(\epsilon_0\vec{E}_{1}+\vec{P}_{e1})=\vec{J}_{e1}+g_{a A B}a\vec{J}_{e_0}
\end{aligned}
\label{AmpereMP2}
\end{equation}
where we define the effective polarization and magnetization as,
\begin{equation}
\begin{aligned}
&\frac{1}{\epsilon_0}\vec{P}_{e1}=g_{a AB} a \vec{E}_{0}-g_{a \gamma\gamma} ca\vec{B}_{0},\\
&\mu_0\vec{M}_{e1}=-\frac{g_{a \gamma\gamma}}{c} a \vec{E}_{0}- g_{a AB}a\vec{B}_{0}. \\
\end{aligned}
\label{MagPol}
\end{equation}

Applying the same vector identities to the magnetic Gauss' and Faraday law we obtain,
\begin{equation}
\begin{aligned}
\vec{\nabla} \cdot(\vec{B}_{1}-\mu_0\vec{M}_{m1})=g_{a B B}a\mu_0c \rho_{e0}
\end{aligned}
\label{MGaussMP2}
\end{equation}
\begin{equation}
\begin{aligned}
\vec{\nabla} \times(\vec{E}_{1}+\frac{1}{\epsilon_0}\vec{P}_{m1})+\partial_t(\vec{B}_{1}-\mu_0\vec{M}_{m1})=-g_{a\mathrm{BB}}ac\mu_0\vec{J}_{e_0}
\end{aligned}
\label{FaradayMP2}
\end{equation}
with the following definitions of effective polarization and magnetization, 
\begin{equation}
\begin{aligned}
&\frac{1}{\epsilon_0}\vec{P}_{m1}=-g_{a AB} a \vec{E}_{0}-g_{a BB}  ca\vec{B}_{0}, \\
&\mu_0\vec{M}_{m1}=-\frac{g_{a BB}}{c} a \vec{E}_{0}+g_{a AB}a \vec{B}_{0}.
\end{aligned}
\label{MagPol2}
\end{equation}
We have reversed the symbols and adopted an opposite sign convention for $\vec{P}_{m1}$ and $\vec{M}_{m1}$ compared to \cite{SokolovMonopole22}, this means we keep the vector $\vec{P}$ as an electrical polarization, and vector $\vec{M}$ as a magnetic polarization (or magnetization), then the subscript refers to whether it is induced from an effective axion like electric ($e$) or magnetic ($m$) charge. The opposite sign convention is used to keep them consistent with how the auxiliary fields ($\vec{D}$ and $\vec{H}$) are defined in matter, which then may be generalised to, 
\begin{equation}
\begin{aligned}
\vec{D}_1&=\epsilon_0\vec{E}_1+\vec{P}_{e1}+\vec{P}_{m1} \\
&=\epsilon_0\vec{E}_1-(g_{a \gamma\gamma} +g_{a BB})a\epsilon_0c\vec{B}_{0},~ \text{and} \\ 
\vec{H}_1&=\frac{1}{\mu_0}\vec{B}_1-\vec{M}_{e1}-\vec{M}_{m1} \\
&=\frac{1}{\mu_0}\vec{B}_1+(g_{a \gamma\gamma} +g_{a BB})a\epsilon_0c \vec{E}_{0}.
\end{aligned}
\label{AuxField}
\end{equation}
Note in this representation of axion modified electrodynamics, both the electric field and magnetic flux densities may have both vector and scalar potentials as dictated by two potential theory \cite{Cabibbo1962,Zwanziger1971,Singleton:1995dp,Singleton96,Keller2018,Rajantie2012,Mignaco2001,Tobar2022b,RHbook2012,Balanis2012}. One can also write these electrodynamic equations in terms of the auxiliary fields. Thus, assuming $\vec{\nabla}a=0$ and combining (\ref{AuxField}) with (\ref{GausMPa})-(\ref{FaradayMPa}), we may write the axion modified electrodynamics as,
\begin{equation}
\begin{aligned}
\vec{\nabla} \cdot\vec{D}_{1}=\rho_{e1},
\end{aligned}
\label{GausAux}
\end{equation}
\begin{equation}
\begin{aligned}
\vec{\nabla} \times\vec{H}_{1}&=\epsilon_0\partial_t{\vec{E}}_{1}+\vec{J}_{e1}+(g_{a A B}\epsilon_0\vec{E}_{0}-g_{a\gamma\gamma}c\epsilon_0\vec{B}_{0})\partial_t{a} \\
&-(g_{a\gamma\gamma}+g_{aBB})a\epsilon_0c\partial_t{\vec{B}_0},
\end{aligned}
\label{AmpereAux}
\end{equation}\begin{equation}
\begin{aligned}
\vec{\nabla} \cdot\vec{H}_{1}=(g_{a \gamma\gamma} +g_{a BB})ac\rho_{e0},
\end{aligned}
\label{MGaussAux}
\end{equation}
\begin{equation}
\begin{aligned}
\frac{1}{\epsilon_0}\vec{\nabla} \times\vec{D}_{1}=-\partial_t{\vec{B}}_{1}+(g_{aAB} \vec{B}_{0}-\frac{g_{aBB}}{c}\vec{E}_{0})\partial_t{a} \\
-\frac{g_{a \gamma\gamma} +g_{a BB}}{c}a\partial_t\vec{E}_{0}-(g_{a\gamma\gamma}+g_{aBB})ac\mu_0\vec{J}_{e0},
\end{aligned}
\label{FaradayAux}
\end{equation}
another form of the modified axion electrodynamics. 

To calculate the sensitivity of experiments to $(g_{a\gamma\gamma},g_{aAB},g_{aBB})$ one can use Poynting theorem, with different choices of vectors considering the electric and magnetic fields and auxiliary fields \cite{Kinsler_2009}. However, similar to what has been shown before \cite{tobar2021abraham}, for resonant and radiative systems the choice of Poynting vector to calculate the sensitivity gives the same first order solution, so in the following analysis we use the simplest form given by equations (\ref{GausMPa})-(\ref{FaradayMPa}) (this is straightforward to show and not included here). 

\section{Phasor Form and Complex Poynting Theorem}

For harmonic solutions, the axion pseudo-scalar $a(t)$ may be written as, $a(t)=\frac{1}{2}\left(\tilde{a} e^{-j \omega_a t}+\tilde{a} ^* e^{j \omega_a t}\right)= \operatorname{Re}\left(\tilde{a} e^{-j \omega_a t}\right)$, and thus, in phasor form, $\tilde{A}=\tilde{a}e^{-j \omega_a t}$ and $\tilde{A}^*=\tilde{a}^*e^{j \omega_a t}$. In contrast, the electric and magnetic fields as well us the electric current are represented as vector-phasors. For example, we set $\vec{E}_1(\vec{r},t)=\frac{1}{2}\left(\mathbf{E}_1(\vec{r})  e^{-j \omega_1 t}+\mathbf{E}_1^*(\vec{r}) e^{j \omega_1 t}\right)=\operatorname{Re}\left[\mathbf{E}_1(\vec{r}) e^{-j \omega_1 t}\right]$, so we define the vector phasor (bold) and its complex conjugate by, $\tilde{\mathbf{E}}_1(\vec{r},t)=\mathbf{E}_1(\vec{r}) e^{-j \omega_1 t}$ and $\tilde{\mathbf{E}}_1^*(\vec{r},t)=\mathbf{E}_1^*(\vec{r}) e^{j \omega_1 t}$, respectively. Following these definitions, the axion modified Ampere's law in (\ref{AmpereMP}), in phasor form becomes,
\begin{equation}
\begin{aligned}
\frac{1}{\mu_0}\vec{\nabla}\times\mathbf{B}_1=\mathbf{J}_{e1}-j\omega_1\epsilon_0\mathbf{E}_{1}&+j\omega_ag_{\alpha \gamma\gamma}\epsilon_0c\tilde{a}\mathbf{B}_0 \\
&-j\omega_ag_{\alpha A B}\epsilon_0 \tilde{a}\mathbf{E}_{0}
\\
\frac{1}{\mu_0}\vec{\nabla}\times\mathbf{B}_1^{*} =\mathbf{J}_{e1}^*+j\omega_1\epsilon_0\mathbf{E}_{1}^{*}&-j\omega_ag_{\alpha \gamma\gamma}\epsilon_0c\tilde{a}^*\mathbf{B}_0^* \\
&+j\omega_ag_{\alpha A B}\epsilon_0\tilde{a}^* \mathbf{E}_{0}^*
\end{aligned}
\label{PhasorAmp}
\end{equation}
while Faraday's law in (\ref{FaradayMP}) becomes,
\begin{equation}
\begin{aligned}
\vec{\nabla}\times \mathbf{E}_1 &=j\omega_1\mathbf{B}_1+j\frac{\omega_ag_{\alpha \mathrm{BB}}}{c}\tilde{a}\mathbf{E}_0-j\omega_ag_{\alpha A B}\tilde{a}\mathbf{B}_{0}
\\
\vec{\nabla}\times \mathbf{E}_1^* &=-j\omega_1\mathbf{B}_1^*-j\frac{\omega_ag_{\alpha \mathrm{BB}}}{c}\tilde{a}^*\mathbf{E}_0^*+j\omega_ag_{\alpha A B}\tilde{a}^*\mathbf{B}_{0}^*
\end{aligned}
\label{PhasorFar}
\end{equation}
To implement Poynting theorem to calculate the sensitivity of a resonant system, we need to calculate the real power flow at resonance as the reactive power is zero at the resonant frequency \cite{tobar2021abraham}. The complex Poynting vector and its complex conjugate are defined by,
\begin{equation}
\mathbf{S}_1=\frac{1}{2\mu_0} \mathbf{E}_1 \times \mathbf{B}_1^{*}~~\text{and}~~\mathbf{S}_1^{*}=\frac{1}{2\mu_0} \mathbf{E}_1^{*} \times \mathbf{B}_1,
\label{AbPv}
\end{equation}
respectively, where $\mathbf{S}_1$ is the complex power density of the harmonic electromagnetic wave or oscillation, with the real part equal to the time averaged power density and the imaginary term equal to the reactive power, which may be inductive (magnetic energy dominates) or capacitive (electrical energy dominates). For resonant systems as analysed in this paper the inductive and capacitive imaginary terms cancel as verified in \cite{tobar2021abraham}, so we only need to consider the real term, however for reactive systems it is the reactive power that dominates, but these systems are not considered in this paper. Unambiguously we may calculate the real part of the Poynting vector by,
\begin{equation}
\begin{aligned}
\operatorname{Re}\left(\mathbf{S}_1\right)=\frac{1}{2}(\mathbf{S}_{1}+\mathbf{S}_{1}^*)
\end{aligned}
\label{ReIm}
\end{equation}
Taking the divergence of Eq. (\ref{ReIm}) we find
\begin{equation}
\begin{aligned}
\nabla\cdot\operatorname{Re}\left(\mathbf{S}_{1}\right)=\frac{1}{2}\nabla\cdot(\mathbf{S}_{1}+\mathbf{S}_{1}^*)\\
\end{aligned}
\label{ReImDivEH}
\end{equation}
The next step is to calculate $\nabla\cdot\mathbf{S}_{1}$ and $\nabla\cdot\mathbf{S}_{1}^*$,
\begin{equation}
\begin{aligned}
&\vec{\nabla}\cdot\mathbf{S}_{1}=\frac{1}{2} \vec{\nabla}\cdot(\mathbf{E}_{1} \times \frac{1}{\mu_0}\mathbf{B}_1^*) =\\
&\frac{1}{2\mu_0}\mathbf{B}_1^* \cdot\vec{\nabla}\times \mathbf{E}_{1}-\mathbf{E}_{1} \cdot\frac{1}{2\mu_0}\vec{\nabla}\times \mathbf{B}_1
\end{aligned}
\label{DivSEH}
\end{equation}
and
\begin{equation}
\begin{aligned}
&\vec{\nabla}\cdot\mathbf{S}_{1}^*=\frac{1}{2} \vec{\nabla}\cdot(\mathbf{E}_{1}^* \times \frac{1}{\mu_0}\mathbf{B}_1) =\\
&\frac{1}{2\mu_0}\mathbf{B}_1 \cdot\vec{\nabla}\times \mathbf{E}_{1}^*-\mathbf{E}_{1}^* \cdot\frac{1}{2\mu_0}\vec{\nabla}\times \mathbf{B}_1.
\end{aligned}
\label{DivSEHIm}
\end{equation}
Substituting Eqs. (\ref{PhasorAmp})  and (\ref{PhasorFar}) into Eqs. (\ref{DivSEH}) and (\ref{DivSEHIm}) leads to,
\begin{equation}
\begin{aligned}
\vec{\nabla}\cdot\mathbf{S}_1&=\frac{j\omega_1\epsilon_0}{2}(c^2\mathbf{B}_1^* \cdot\mathbf{B}_1-\mathbf{E}_1 \cdot\mathbf{E}_{1}^{*})-\frac{1}{2}\mathbf{E}_1 \cdot\mathbf{J}_{e1}^* \\
&-\frac{j\omega_a\epsilon_0g_{a A B}}{2}(c^2\mathbf{B}_1^* \cdot\tilde{a}\mathbf{B}_{0}+\mathbf{E}_1 \cdot\tilde{a}^* \mathbf{E}_{0}^*) \\
&+\frac{j\omega_a\epsilon_0c}{2}(g_{a BB}\mathbf{B}_1^* \cdot\tilde{a}\mathbf{E}_0
+g_{a\gamma\gamma}\mathbf{E}_1 \cdot\tilde{a}^*\mathbf{B}_0^*)
\end{aligned}
\label{divReExEH}
\end{equation}
\begin{equation}
\begin{aligned}
\vec{\nabla}\cdot\mathbf{S}_{1}^*&=\frac{j\omega_1\epsilon_0}{2}(\mathbf{E}_1 \cdot\mathbf{E}_{1}^{*}-c^2\mathbf{B}_1^* \cdot\mathbf{B}_1)-\frac{1}{2}\mathbf{E}_1^* \cdot\mathbf{J}_{e1}\\
&+\frac{j\omega_a\epsilon_0g_{a A B}}{2}(c^2\mathbf{B}_1 \cdot\tilde{a}^*\mathbf{B}_{0}^*+\mathbf{E}_1^* \cdot\tilde{a} \mathbf{E}_{0}) \\
&-\frac{j\omega_a\epsilon_0c}{2}(g_{a BB}\mathbf{B}_1 \cdot\tilde{a}^*\mathbf{E}_0^*+g_{a\gamma\gamma}\mathbf{E}_1^* \cdot\tilde{a}\mathbf{B}_0).
\end{aligned}
\label{divImExEH}
\end{equation}
Now by substituting (\ref{divReExEH}) and (\ref{divImExEH}) into (\ref{ReImDivEH}), we find,
\begin{equation}
\begin{aligned}
&\nabla\cdot\operatorname{Re}\left(\mathbf{S}_{1}\right)= -\frac{1}{4}(\mathbf{E}_1 \cdot\mathbf{J}_{e1}^*-\mathbf{E}_1^* \cdot\mathbf{J}_{e1})\\
&+\frac{j\omega_a\epsilon_0c^2g_{a A B}}{4}(\mathbf{B}_1 \cdot\tilde{a}^*\mathbf{B}_{0}^*-\mathbf{B}_1^* \cdot\tilde{a}\mathbf{B}_{0}) \\
&+\frac{j\omega_a\epsilon_0g_{a A B}}{4}(\mathbf{E}_1^* \cdot\tilde{a} \mathbf{E}_{0}-\mathbf{E}_1 \cdot\tilde{a}^* \mathbf{E}_{0}^*) \\
&+\frac{j\omega_a\epsilon_0cg_{a BB}}{4}(\mathbf{B}_1^* \cdot\tilde{a}\mathbf{E}_0-\mathbf{B}_1 \cdot\tilde{a}^*\mathbf{E}_0^*) \\
&+\frac{j\omega_a\epsilon_0cg_{a\gamma\gamma}}{4}(\mathbf{E}_1 \cdot\tilde{a}^*\mathbf{B}_0^*
-\mathbf{E}_1^* \cdot\tilde{a}\mathbf{B}_0),
\end{aligned}
\end{equation}
then applying the divergence theorem, we obtain
\begin{equation}
\begin{aligned}
&\oint\operatorname{Re}\left(\mathbf{S}_{1}\right)\cdot \hat{n}ds=\int\Big(-\frac{1}{4}(\mathbf{E}_1 \cdot\mathbf{J}_{e1}^*-\mathbf{E}_1^* \cdot\mathbf{J}_{e1})\\
&+\frac{j\omega_a\epsilon_0c^2g_{a A B}}{4}(\mathbf{B}_1 \cdot\tilde{a}^*\mathbf{B}_{0}^*-\mathbf{B}_1^* \cdot\tilde{a}\mathbf{B}_{0}) \\
&+\frac{j\omega_a\epsilon_0g_{a A B}}{4}(\mathbf{E}_1^* \cdot\tilde{a} \mathbf{E}_{0}-\mathbf{E}_1 \cdot\tilde{a}^* \mathbf{E}_{0}^*) \\
&+\frac{j\omega_a\epsilon_0cg_{a BB}}{4}(\mathbf{B}_1^* \cdot\tilde{a}\mathbf{E}_0-\mathbf{B}_1 \cdot\tilde{a}^*\mathbf{E}_0^*) \\
&+\frac{j\omega_a\epsilon_0cg_{a\gamma\gamma}}{4}(\mathbf{E}_1 \cdot\tilde{a}^*\mathbf{B}_0^*
-\mathbf{E}_1^* \cdot\tilde{a}\mathbf{B}_0)\Big)~dV.
\end{aligned}
\label{PoyntTh}
\end{equation}
This equation may be used to calculate the expected power in the photon-axion energy conversion in a resonant system.

\section{Sensitivity of Axion Resonant Haloscopes under DC Magnetic Fields }

In this section we calculate the sensitivity of conventional Sikivie-type resonant axion haloscope \cite{hagmann1990,Hagmann1990b,Bradley2003,ADMXaxions2010,ADMX2011,Braine2020,Bartram2021,McAllister2017,Quiskamp2022,universe7070236,10.1093/ptep/ptab051,Devlin2021,Kwon2021,Backes:2021wd,IWAZAKI2021115298,Chigusa:2021vh,PhysRevD.103.096001,Alvarez-Melcon:2021aa,Alesini22}. If we assume that the background degree of freedom is only excited by a DC magnetic field, then background field equations (\ref{Bground}) become,
\begin{equation}
\begin{aligned}
&\vec{\nabla}\times \vec{B}_0=\mu_0\vec{J}_{e_0}~~~~~\vec{\nabla}\cdot \vec{B}_0=0~~~~~~\vec{E}_0=0,
\end{aligned}
\label{BgroundDCB}
\end{equation}
Since the phase of the photon leaving the cavity is not an observable, we may arbitrarily set the axion phase, setting $a_0=\tilde{a}=\tilde{a}^*=\sqrt{2}\langle a_0\rangle$, (\ref{PoyntTh}) becomes,
\begin{equation}
\begin{aligned}
&\oint\operatorname{Re}\left(\mathbf{S}_{1}\right)\cdot \hat{n}ds=\int\Big(-\frac{1}{4}(\mathbf{E}_1 \cdot\mathbf{J}_{e1}^*-\mathbf{E}_1^* \cdot\mathbf{J}_{e1})+\\
&\frac{j\omega_a\epsilon_0\langle a_0\rangle}{2\sqrt{2}}\vec{B}_0\cdot\big(c^2g_{a A B}(\mathbf{B}_1-\mathbf{B}_1^*)+g_{a\gamma\gamma}(\mathbf{E}_1-\mathbf{E}_1^*)\big)\Big)~dV
\end{aligned}
\label{PoyntThBDC2}
\end{equation}
First thing we notice is that it is quite clear that a haloscope with a DC magnetic field is not sensitive to $g_{aBB}$. 

In the following, we assume a resonant cavity of volume, $V_1$, with resonant modes of stored electromagnetic energy, $U_1$, given by,
\begin{equation}
\begin{aligned}
U_1=\frac{\epsilon_0}{4}\int\mathbf{E}_1 \cdot\mathbf{E}_1^*dV+\frac{1}{4\mu_0}\int\mathbf{B}_1 \cdot\mathbf{B}_1^*dV,
\end{aligned}
\label{StEngB}
\end{equation}
where the first term is the electric stored energy and the second term is the magnetic. The electric and magnetic energy must be equal, thus we may also write $U_1=\frac{\epsilon_0}{2}\int\mathbf{E}_1 \cdot\mathbf{E}_1^*dV=\frac{1}{2\mu_0}\int\mathbf{B}_1 \cdot\mathbf{B}_1^*dV$. The first term on the right-hand side of (\ref{PoyntThBDC2}) represents the power dissipated in the resonator. For example, the electric field on the surface of the cavity where the power is dissipated is given by $\mathbf{E}_1= R_s\mathbf{K}_1$, where $R_s$ is the surface resistance and $\mathbf{K}_1$ the surface current. Substituting these values into $P_d=\frac{1}{4}\int(\mathbf{E}_1 \cdot\mathbf{J}_{e1}^*-\mathbf{E}_1^* \cdot\mathbf{J}_{e1})~dV$, the integral collapses to the surface integral, $P_d=\frac{R_S}{2}\oint\mathbf{K}_1\cdot\mathbf{K}_{1}^*ds=\frac{R_S}{2}\oint\mathbf{H}_{1}\cdot\mathbf{H}_{1}^*ds=\frac{\omega_1U_1}{Q_1}=\frac{\omega_1\epsilon_0c^2}{2Q_1}\int\mathbf{B}_1 \cdot\mathbf{B}_1^*~dV=\frac{\omega_1\epsilon_0}{2Q_1}\int\mathbf{E}_1 \cdot\mathbf{E}_1^*~dV$. From this relation the well-known geometry factor, which relates the cavity surface resistance to the cavity Q may be derived,
\begin{equation}
\begin{aligned}
G_1=R_s Q_1=\frac{\omega_1\mu_0\int\mathbf{H}_1 \cdot\mathbf{H}_1^*~dV}{\oint\mathbf{H}_{1}\cdot\mathbf{H}_{1}^*ds}.
\end{aligned}
\label{Geom}
\end{equation}
In this case the dissipative components of $\mathbf{E}_1$ and $\mathbf{B}_1$ fields are attenuated in the cavity walls in a similar way and are effectively in phase, and we may write the fields as complex with respect to a loss tangent $(\tan\delta_1\sim\frac{1}{Q_1})$, so $\mathbf{B}_{1}\approx(1-j\frac{1}{Q_1})\operatorname{Re}(\mathbf{B}_{1})$ and $\mathbf{E}_{1}\approx(1-j\frac{1}{Q_1})\operatorname{Re}(\mathbf{E}_{1})$, where the real terms relates to the oscillating fields over the volume and the imaginary terms relate to dissipative fields, and thus $\mathbf{E}_1-\mathbf{E}_1^*=-j\frac{2}{Q_1}\operatorname{Re}(\mathbf{E}_{1}$). Furthermore, on resonance and in steady state we may assume $\oint\operatorname{Re}\left(\mathbf{S}_{1}\right)\cdot \hat{n}ds=0$, meaning there is no external energy inputted at the resonant mode frequency, so the axion generated signal power, $P_{s1}$ in the resonant mode is equal to the dissipated power, $P_d$, so $P_{s1}$ in equation (\ref{PoyntThBDC2}) may be identified as,
\begin{equation}
\begin{aligned}
P_{s1}=P_d&=\frac{\omega_1U_1}{Q_1}=g_{a\gamma\gamma}\frac{\omega_a\epsilon_0\langle a_0\rangle}{\sqrt{2}Q_1}\int \vec{B}_0\cdot\operatorname{Re}(\mathbf{E}_1)~dV\\
&+g_{a A B}\frac{\omega_a\epsilon_0\langle a_0\rangle c}{\sqrt{2}Q_1}\int \vec{B}_0\cdot\operatorname{Re}(\mathbf{B}_1)~dV.
\end{aligned}
\label{PVDCB}
\end{equation}
Combining (\ref{StEngB}) with (\ref{PVDCB}) we may obtain,
\begin{equation}
\begin{aligned}
\sqrt{U_1}=\frac{\omega_a\sqrt{\epsilon_0}\langle a_0\rangle}{\omega_1}&\Big(g_{a\gamma\gamma}\frac{\int\vec{B}_0\cdot\operatorname{Re}(\mathbf{E}_1)~dV}{\sqrt{\int\mathbf{E}_1 \cdot\mathbf{E}_1^*~dV}} \\
&+g_{aAB}\frac{\int \vec{B}_0\cdot\operatorname{Re}(\mathbf{B}_1)~dV}{\sqrt{\int\mathbf{B}_1\cdot\mathbf{B}_1^*~dV}}\Big)
\end{aligned}
\label{SqrtStEng}
\end{equation}
Now defining the following form factors of the haloscope,
\begin{equation}
\begin{aligned}
C_{1a\gamma\gamma}&=\frac{(\int \vec{B}_0\cdot\operatorname{Re}(\mathbf{E}_{1})dV)^2}{B_0^2V_1\int\mathbf{E}_1\cdot\mathbf{E}_{1}^*dV} \\
C_{1AB}&=\frac{(\int \vec{B}_0\cdot\operatorname{Re}(\mathbf{B}_{1})dV)^2}{B_0^2V_1\int\mathbf{B}_1\cdot\mathbf{B}_{1}^*dV},
\end{aligned}
\label{FormFac}
\end{equation}
 and setting $\langle a_{0}\rangle^2=\frac{\rho_{a}}{c} \frac{\hbar^2}{m_{a}^2}=\frac{\rho_{a}c^3}{\omega_a^2}$, where $\rho_{a}$ is the axion dark matter density, the signal gained at the output of the haloscope becomes,
\begin{equation}
\begin{aligned}
&\sqrt{P_{1}}=\sqrt{\omega_aQ_1U_1} \\
&=(g_{a\gamma\gamma}\sqrt{C_{1a\gamma\gamma}}+g_{aAB}\sqrt{C_{1aAB}})\langle a_0\rangle cB_0\sqrt{\omega_aQ_1\epsilon_0V_1}\\
&=(g_{a\gamma\gamma}\sqrt{C_{1a\gamma\gamma}}+g_{aAB}\sqrt{C_{1aAB}})B_0\sqrt{\frac{\rho_{a}Q_1\epsilon_0c^5V_1}{\omega_a}},
\end{aligned}
\label{FormFac2}
\end{equation}

Considering the modes in a cylindrical cavity, with the z-axis aligned with a DC magnetic field, so $\vec{B}_0=B_z\hat{z}$, then $\int \mathbf{B}_1\cdot \hat{z}~dV =0$ for all modes if $B_z$ is constant, and the calculation is consistent with what has been derived previously \cite{Sikivie1984,Sikivie2021,McAllisterFormFactor}, equivalent to the well-known sensitivity equation of a Sikivie-type axion haloscope, which to first order is not sensitive to $g_{a A B}$ or $g_{a BB}$ and is only sensitive to $g_{a\gamma\gamma}$.

To gain sensitivity to  $g_{a A B}$ with a DC haloscope, one can apply a nonuniform DC magnetic field with other vector components or a DC electric field (see next section), in a similar way to scalar-field dark matter experiments recently proposed in \cite{Samsonov2022}, which set limits on the dilaton scalar coupling parameter, $g_{\phi\gamma\gamma}$. It turns out that the limits set on $g_{\phi\gamma\gamma}$ in \cite{Samsonov2022} are equivalent to limits on the axion-photon parameter, $g_{a A B}$, so in effect this paper also sets limits on $g_{a A B}$.

\section{Sensitivity of Axion Resonant Haloscopes under DC Electric Fields}

If we assume that the background degree of freedom is only excited by a DC electric field, then background field equations (\ref{Bground}) become,
\begin{equation}
\begin{aligned}
&\vec{\nabla}\times \vec{E}_0=0~~~~\epsilon_0\vec{\nabla}\cdot \vec{E}_0=\rho_{e_0}~~~~~~~\vec{B}_0=0,
\end{aligned}
\label{BgroundDCB2}
\end{equation}
Since the phase of the photon leaving the cavity is not an observable, we may arbitrarily set the axion phase, setting $\tilde{a}=\tilde{a}^*=-a_0=-\sqrt{2}\langle a_0\rangle$ (\ref{PoyntTh}) becomes,
\begin{equation}
\begin{aligned}
&\oint\operatorname{Re}\left(\mathbf{S}_{1}\right)\cdot \hat{n}ds=\int\Big(-\frac{1}{4}(\mathbf{E}_1 \cdot\mathbf{J}_{e1}^*-\mathbf{E}_1^* \cdot\mathbf{J}_{e1})\\
&+\frac{j\omega_a\epsilon_0\langle a_0\rangle g_{a A B}}{2\sqrt{2}}\vec{E}_{0}\cdot(\mathbf{E}_1-\mathbf{E}_1^*) \\
&+\frac{j\omega_a\epsilon_0\langle a_0\rangle cg_{a BB}}{2\sqrt{2}}\vec{E}_0\cdot(\mathbf{B}_1-\mathbf{B}_1^*)\Big)~dV.
\end{aligned}
\label{PoyntThEDC}
\end{equation}
Now defining the following form factors of the haloscope,
\begin{equation}
\begin{aligned}
C_{1ABm}&=\frac{(\int \vec{E}_0\cdot\operatorname{Re}(\mathbf{E}_{1})dV)^2}{E_0^2V_1\int\mathbf{E}_1\cdot\mathbf{E}_{1}^*dV} \\
C_{1BB}&=\frac{(\int \vec{E}_0\cdot\operatorname{Re}(\mathbf{B}_{1})dV)^2}{E_0^2V_1\int\mathbf{B}_1\cdot\mathbf{B}_{1}^*dV},
\end{aligned}
\label{FormFac3}
\end{equation}
the signal gained at the output of the haloscope becomes,
\begin{equation}
\begin{aligned}
&\sqrt{P_{1}}=\sqrt{\omega_aQ_1U_1} \\
&=(g_{aBB}\sqrt{C_{1aBB}}+g_{aAB}\sqrt{C_{1aABm}})\langle a_0\rangle E_0\sqrt{\omega_aQ_1\epsilon_0V_1}\\
&=(g_{aBB}\sqrt{C_{1aBB}}+g_{aAB}\sqrt{C_{1aABm}})E_0\sqrt{\frac{\rho_{a}Q_1\epsilon_0c^3V_1}{\omega_a}},
\end{aligned}
\label{FormFac4}
\end{equation}

Thus, by applying a DC electric field to a cavity resonator, the experiment becomes sensitive to $g_{aBB}$ and $g_{aAB}$, where the latter is sensitive to $TM_{0,n,0}$ modes in a cylindrical cavity resonator if an electric field is applied along the cylinder axis. However, to attain sensitivity to $g_{aBB}$ a more complicated DC electric field is required. It is a much harder experiment to supply a large DC electric field across a high-Q tunable cavity, even though the QCD axion is supposed to have a larger coupling to $g_{aBB}$ than $g_{a\gamma\gamma}$ \cite{SokolovMonopole22}. Thus, the resonant cavity technique might not be the optimum way to make use of this larger coupling, as it is much easier to configure an experiment with a large DC magnetic field in comparison to a large DC electric field.

\section{Sensitivity of Upconversion Resonant Haloscopes}

The upconversion technique utilises two modes of a resonant cavity, a readout mode (subscript 1) and a background mode (subscript 0), and will up converts a putative axion signal of mass equivalent to the difference frequency between the two modes, so $\omega_a\sim|\omega_1-\omega_0|$, where $\omega_a<<\omega_1$. This technique was first proposed in~\cite{Goryachev2019,Thomson:2021wk} and experimentally demonstrated in \cite{Cat21}, and showed that a dark matter axion will perturb the frequency (or phase) and amplitude (or power) of the readout mode. The former we call the ``frequency technique'' and the later the ``power technique''. The power technique was also proposed in \cite{berlin2020axion,Lasenby2020b,ABerlin2021,Lasenby2020}, and later performed in~\cite{Cat23}. In this work we calculate the sensitivity of this experiment to the three axion coupling parameters $(g_{a\gamma\gamma},g_{aAB},g_{aBB})$.

\subsubsection{Power Technique}

Here we use the Poynting vector equation (\ref{PoyntTh}) to derive the sensitivity of the power technique, where the background field will mix with the axion to generate power at the readout mode frequency.  Since the phase of the photon leaving the cavity is not our observable, we may arbitrarily set the axion phase, setting $a_0=\tilde{a}=\tilde{a}^*=\sqrt{2}\langle a_0\rangle$, equation (\ref{PoyntTh}) becomes,
\begin{equation}
\begin{aligned}
&\oint\operatorname{Re}\left(\mathbf{S}_{1}\right)\cdot \hat{n}ds=\int\Big(-\frac{1}{4}(\mathbf{E}_1 \cdot\mathbf{J}_{e1}^*-\mathbf{E}_1^* \cdot\mathbf{J}_{e1})\\
&+\frac{j\omega_a\epsilon_0cg_{a BB}\sqrt{2}\langle a_0\rangle}{4}(\mathbf{B}_1^* \cdot\mathbf{E}_0-\mathbf{B}_1 \cdot\mathbf{E}_0^*) \\
&+\frac{j\omega_a\epsilon_0cg_{a\gamma\gamma}\sqrt{2}\langle a_0\rangle}{4}(\mathbf{E}_1 \cdot\mathbf{B}_0^*-\mathbf{E}_1^* \cdot\mathbf{B}_0)\Big)~dV,
\end{aligned}
\label{PoyntAC}
\end{equation}
where we can identify the power generated by the axion as,
\begin{equation}
\begin{aligned}
P_{s1}&=\frac{j\omega_a\epsilon_0cg_{a BB}\sqrt{2}\langle a_0\rangle}{4}\int\Big((\mathbf{B}_1^* \cdot\mathbf{E}_0-\mathbf{B}_1 \cdot\mathbf{E}_0^*) \\
&+\frac{j\omega_a\epsilon_0cg_{a\gamma\gamma}\sqrt{2}\langle a_0\rangle}{4}(\mathbf{E}_1 \cdot\mathbf{B}_0^*-\mathbf{E}_1^* \cdot\mathbf{B}_0)\Big)~dV.
\end{aligned}
\label{PVDM1}
\end{equation}
Note, the sensitivity coefficients to $g_{a AB}$ drop out, as in the lossless limit they are identically zero.

Ignoring losses in the background fields we individually consider them real for both $\mathbf{E}_0$ and $\mathbf{B}_0$. \begin{equation}
\begin{aligned}
P_{s1}=\frac{\omega_a\epsilon_0c\langle a_0\rangle}{\sqrt{2}Q_1}&\int\Big(g_{a\gamma\gamma}\operatorname{Re}(\mathbf{E}_1)\cdot\operatorname{Re}(\mathbf{B}_0) \\
&-g_{a BB}\operatorname{Re}(\mathbf{B}_1)\cdot\operatorname{Re}(\mathbf{E}_0)\Big)~dV.
\end{aligned}
\label{PVDM2}
\end{equation}
Equating $P_{s1}$ in (\ref{PVDM2}) to $P_d=\frac{\omega_1U_1}{Q_1}$, we can write the stored energy as,
\begin{equation}
\begin{aligned}
U_1=\frac{\omega_a\epsilon_0\langle a_0\rangle}{\sqrt{2}\omega_1}\Big(g_{a\gamma\gamma}\int\operatorname{Re}(\mathbf{E}_1)\cdot\operatorname{Re}(c\mathbf{B}_0)~dV \\
-g_{a BB}\int\operatorname{Re}(c\mathbf{B}_1)\cdot\operatorname{Re}(\mathbf{E}_0)~dV\Big).
\end{aligned}
\label{StEngEB}
\end{equation}
Since $U_1=\frac{\epsilon_0}{2}\int\mathbf{E}_1 \cdot\mathbf{E}_1^*dV$, we obtain,
\begin{equation}
\begin{aligned}
\sqrt{U_1}=\frac{\omega_a\epsilon_0\langle a_0\rangle}{\sqrt{2}\omega_1}&\Big(g_{a\gamma\gamma}\frac{\int\operatorname{Re}(\mathbf{E}_1)\cdot\operatorname{Re}(c\mathbf{B}_0)~dV}{\sqrt{\frac{\epsilon_0}{2}\int\mathbf{E}_1 \cdot\mathbf{E}_1^*~dV}} \\
&-g_{a BB}\frac{\int\operatorname{Re}(c\mathbf{B}_1)\cdot\operatorname{Re}(\mathbf{E}_0)~dV}{\sqrt{\frac{\epsilon_0}{2}\int\mathbf{E}_1\cdot\mathbf{E}_1^*~dV}}\Big)
\end{aligned}
\label{SqrtStEngEB}
\end{equation}
Then defining the unit vectors, so $c\mathbf{B}_0=E_{00}\mathbf{b}_0$ and $\mathbf{E}_0=E_{00}\mathbf{e}_0$, and $c\mathbf{B}_1=E_{01}\mathbf{b}_1$ and $\mathbf{E}_1=E_{01}\mathbf{e}_1$ then (\ref{SqrtStEngEB}) becomes,
\begin{equation}
\begin{aligned}
\sqrt{U_1}=\frac{\omega_a\sqrt{\epsilon_0}\sqrt{V}E_{00}\langle a_0\rangle}{\omega_1}
&\Big(g_{a\gamma\gamma}\frac{\frac{1}{V}\int\mathbf{e}_1\cdot\mathbf{b}_0~dV}{\sqrt{\frac{1}{V}\int\mathbf{e}_1 \cdot\mathbf{e}_1^*~dV}}\\
&-g_{a BB}\frac{\frac{1}{V}\int\mathbf{b}_1\cdot\mathbf{e}_0~dV}{\sqrt{\frac{1}{V}\int\mathbf{e}_1 \cdot\mathbf{e}_1^*~dV}}\Big) \\
=\langle a_0\rangle\frac{\omega_a}{\omega_1}\sqrt{\frac{2P_{c0}}{\omega_0}}&\Big(g_{a\gamma\gamma}\xi_{10}-g_{a BB}\xi_{01}\Big),
\end{aligned}
\label{U1}
\end{equation}
where the overlap functions are defined by \cite{Goryachev2019}, 
\begin{equation}
\xi_{10}=\frac{1}{V}\int\mathbf{e}_1\cdot\mathbf{b}_0dV~~\text{and}~~
 \xi_{01}= \frac{1}{V}\int\mathbf{e}_0\cdot\mathbf{b}_1dV, 
\end{equation}
where the square of the overlap functions are analagous to the form factors in the previous sections. 

Here $E_{00}=\sqrt{\frac{2P_{c0}}{\omega_0\epsilon_0V}}$ 
where $P_{c0}$ is the circulating power of the background mode over the cavity volume $V$,which is related to the incident power, $P_{0inc}$ by,
\begin{equation}
P_{c0}=\frac{4 \beta_{0} Q_{L0}}{(\beta_0+1)^2}P_{0inc},
\end{equation}
where $\beta_0$ is the background mode coupling to the cavity and $Q_{L0}$ the mode loaded quality factor. Now, we can determine the square root power in the coupling circuit of the readout mode to be,
\begin{equation}
\sqrt{P_{1out}}=\frac{\sqrt{\omega_1Q_{L1}U_1}\sqrt{\beta_1}}{\sqrt{1+\beta_1}\sqrt{1+4Q^2_{L1}\big(\frac{\delta\omega_a}{\omega_1}\big)^2}},
\label{Pro}
\end{equation}
where we define $\delta\omega_a=\omega_1+\omega_a-\omega_0$, so when $\delta\omega_a=0$ then $\omega_a=\omega_0-\omega_1$ and the axion induced power is upconverted to the frequency, $\omega_1$. Thus, $\delta\omega_a$ defines the detuning of the induced power with respect to the readout mode frequency.
Combining (\ref{U1})-(\ref{Pro}), we obtain
\begin{equation}
\begin{aligned}
\sqrt{P_{1out}}=\mathcal{K}p_{a\gamma\gamma}~g_{a \gamma\gamma}\left\langle a_{0}\right\rangle+\mathcal{K}p_{aBB}~g_{a BB}\left\langle a_{0}\right\rangle,
\end{aligned}
\label{PowerSens}
\end{equation}
where
\begin{equation}
\begin{aligned}
&\mathcal{K}p_{a\gamma\gamma}=\frac{\xi_{10}2\sqrt{2}\omega_a\sqrt{\beta_{0}Q_{L0}\beta_1Q_{L1}P_{0inc}}}{\sqrt{\omega_1\omega_0}\sqrt{1+\beta_1}(\beta_0+1)\sqrt{1+4Q^2_{L1}\big(\frac{\delta\omega_a}{\omega_1}\big)^2}}, \\
&\mathcal{K}p_{aBB}=-\frac{\xi_{01}2\sqrt{2}\omega_a\sqrt{\beta_{0}Q_{L0}\beta_1Q_{L1}P_{0inc}}}{\sqrt{\omega_1\omega_0}\sqrt{1+\beta_1}(\beta_0+1)\sqrt{1+4Q^2_{L1}\big(\frac{\delta\omega_a}{\omega_1}\big)^2}},
\end{aligned}
\label{PowerSens2}
\end{equation}
which are the axion gain coefficients in units of square root power. Equation (\ref{PowerSens}) can be used to calculate the sensitivity for the power technique, and is sensitive to the effective monopole coupling term $g_{aBB}$ unlike the Sikivie-type detectors that utilises a DC $\vec{B}$ field.

To calculate the signal to noise ratio (SNR) for virialised axion dark matter from the galactic halo, we take into account that it presents as a narrow band noise source with a line width of a part in $10^6$. In SI units we may relate the axion amplitude to the background  dark matter density in the galactic halo, $\rho_a$, by $\left\langle a_{0}\right\rangle=\frac{\sqrt{\rho_{a} c^{3}}}{\omega_{a}}$. For $i= a\gamma\gamma$ or $i= aBB$, then limits on the axion couplings can be undertaken independently by calculating,
\begin{equation}\label{SNR}
SNRp_{i}=g_i \frac{|\mathcal{K}p_{i}|}{\omega_{a}\sqrt{P_N}}\sqrt{\rho_{a} c^{3}}\left(\frac{t}{\Delta f_a}\right)^{\frac{1}{4}},
\end{equation}
where, $P_N$ (Watts/Hz) is the noise power competing with the axion signal and $\Delta f_a$ is the axion bandwidth in Hz, where $\Delta f_a=\frac{f_a}{10^6}$ for virialised dark matter. This assumes the measurement time, $t$ is greater than the axion coherence time so that $t>\Delta f_a^{-1}$. For measurement times of $t<\frac{10^6}{ f_{a}}$ we substitute $\left(\frac{10^6t}{ f_{a}}\right)^{\frac{1}{4}} \rightarrow t^{\frac{1}{2}}$. The noise power in such experiments is dominated by thermal noise in the readout mode resonator of effective temperature, $T_1$ and the noise temperature of the first amplifier, $T_{amp}$ after the readout mode, and is given by \cite{HARTNETT2011,parker2013b},
\begin{equation}
\begin{aligned}
P_N\sim\frac{4 \beta_1 }{(\beta_1 +1)^2 \left(1+4 Q_{L1}^2\big(\frac{\delta\omega_a}{\omega_1}\big)^2\right)} \frac{k_BT_{1}}{2}+\frac{k_BT_{amp}}{2}.
\end{aligned}
\label{PN}
\end{equation}
In the case $\beta_1\sim1$ and $\delta\omega_a\sim0$ then $P_N\sim\frac{k_B(T_1+T_{amp})}{2}$, and assuming $\beta_0\sim1$ the signal to noise ratios become,
\begin{equation}
\begin{aligned}
&SNR_{a\gamma\gamma}\sim g_{a\gamma\gamma}
|\xi_{10}|\sqrt{\frac{2Q_{L0}Q_{L1}P_{0inc}\rho_a c^3}{\omega_1\omega_0k_B(T_1+T_{amp})}}
\left(\frac{10^6t}{ f_{a}}\right)^{\frac{1}{4}}, \\
&SNR_{aBB}\sim g_{aBB}
|\xi_{01}|\sqrt{\frac{2Q_{L0}Q_{L1}P_{0inc}\rho_a c^3}{\omega_1\omega_0k_B(T_1+T_{amp})}}
\left(\frac{10^6t}{ f_{a}}\right)^{\frac{1}{4}}.
\end{aligned}
\label{SNRa}
\end{equation} 

\begingroup
\allowdisplaybreaks
\subsubsection{Resonant Haloscope Frequency Shift from Perturbation Analysis}

As pointed out previously, there is no first order frequency shift for DC Sikivie haloscopes as they are only second order sensitive to frequency shifts \cite{Goryachev2019}. However, when two AC modes are excited, the situation is different as the DC and AC haloscopes belong to different classes of detectors. Since virtual photons or static fields carry no phase, the DC haloscope belongs to the class of phase insensitive systems. In contrast, the AC scheme relies on a pump signal carrying relative phases to the readout signal and axion field, and since this occurs in a resonant cavity, phase shifts are converted to frequency shifts. This is analogous to existing amplifiers that can be grouped into DC (phase insensitive) amplifiers, where energy is drawn from a static power supply, and parametric (phase sensitive) amplifiers, where energy comes from oscillating fields.

To calculate frequency shifts here we adapt the perturbation theory technique to axion modified electrodynamics as opposed to the quantum optics technique used in the past \cite{Goryachev2019,Thomson:2021wk}, which gives the same result. We consider the perturbed readout resonator mode fields, $(\mathbf{E}_{1}^\prime,\mathbf{B}_{1}^\prime)$ and frequency, $\omega_1^\prime$, due to the mixing of the axion and the background pump mode, where the unperturbed modes and frequency are given by Ampere's law and Faraday's law in standard electrodynamics,
\begin{equation}
\begin{aligned}
&\frac{1}{\mu_0}\vec{\nabla}\times\mathbf{B}_1=\mathbf{J}_{e1}-j\omega_1\epsilon_0\mathbf{E}_{1};~~~\vec{\nabla}\times \mathbf{E}_1 =j\omega_1\mathbf{B}_1 \\
&\frac{1}{\mu_0}\vec{\nabla}\times\mathbf{B}_1^{*} =\mathbf{J}_{e1}^*+j\omega_1\epsilon_0\mathbf{E}_{1}^{*}; ~~~\vec{\nabla}\times \mathbf{E}_1^* =-j\omega_1\mathbf{B}_1^*
\end{aligned}
\label{AmpFar}
\end{equation}
and the perturbed  Ampere's law and Faraday's law, are derived from (\ref{PhasorAmp}) and (\ref{PhasorFar}), and written as
\begin{equation}
\begin{aligned}
\frac{1}{\mu_0}\vec{\nabla}\times\mathbf{B}_1^\prime &=\mathbf{J}_{e1}^\prime-j\omega_1^\prime\epsilon_0\mathbf{E}_{1}^\prime \\ 
&+j\omega_ag_{a\gamma\gamma}\epsilon_0c\tilde{a}\mathbf{B}_0 -j\omega_ag_{a A B}\epsilon_0 \tilde{a}\mathbf{E}_{0} \\
\frac{1}{\mu_0}\vec{\nabla}\times\mathbf{B}_1^{*\prime} &=\mathbf{J}_{e1}^{*\prime}+j\omega_1^\prime\epsilon_0\mathbf{E}_{1}^{*\prime} \\
&-j\omega_ag_{a\gamma\gamma}\epsilon_0c\tilde{a}^*\mathbf{B}_0^{*}+j\omega_ag_{a A B}\epsilon_0\tilde{a}^* \mathbf{E}_{0}^{*}\\
\end{aligned}
\label{AmpDash}
\end{equation}
and
\begin{equation}
\begin{aligned}
\vec{\nabla}\times \mathbf{E}_1^\prime &=j\omega_1^\prime\mathbf{B}_1^\prime \\
&+j\frac{\omega_ag_{a BB}}{c}\tilde{a}\mathbf{E}_0-j\omega_ag_{a A B}\tilde{a}\mathbf{B}_{0} \\
\vec{\nabla}\times \mathbf{E}_1^{*\prime} &=-j\omega_1^\prime\mathbf{B}_1^{*\prime} \\
&-j\frac{\omega_ag_{a BB}}{c}\tilde{a}^*\mathbf{E}_0^*+j\omega_ag_{a A B}\tilde{a}^*\mathbf{B}_{0}^*
\end{aligned}
\label{FarDash}
\end{equation}
Then we can implement the following integral,
\begin{equation}
\begin{aligned}
&\frac{1}{\mu_0}\int_V\nabla\cdot\left(\mathbf{E}_1^{*}\times \mathbf{B}_1^{\prime}+\mathbf{E}_1^{\prime} \times \mathbf{B}_1^{*}\right)dV= \\
&\frac{1}{\mu_0}\int_V(\mathbf{B}_1^{\prime}\cdot\nabla\times\mathbf{E}_1^{*}-\mathbf{E}_1^{*}\cdot\nabla\times\mathbf{B}_1^{\prime}+\mathbf{B}_1^{*}\cdot\nabla\times\mathbf{E}_1^{\prime} \\
&-\mathbf{E}_1^{\prime}\cdot\nabla\times\mathbf{B}_1^{*})\, dV \, .
\end{aligned}
\label{PertPont}
\end{equation}
Substituting (\ref{AmpFar}), (\ref{AmpDash}) and (\ref{FarDash}) into (\ref{PertPont}), one obtains,
\begin{eqnarray}
&&\frac{1}{\mu_0}\int_V\nabla\cdot\left(\mathbf{E}_1^{*}\times \mathbf{B}_1^{\prime}+\mathbf{E}_1^{\prime}\times\mathbf{B}_1^{*}\right)dV~= \nonumber \\
&&j\delta\omega_1\int_V (\epsilon_0\mathbf{E}_1^{*}\cdot\mathbf{E}_1^{\prime}+\frac{1}{\mu_0}\mathbf{B}_1^{*}\cdot\mathbf{B}_1^{\prime})dV \nonumber \\
&&-j\omega_a\epsilon_0c\int_V(g_{a\gamma\gamma}\mathbf{E}_1^{*}\cdot\tilde{a}\mathbf{B}_0-g_{aBB}\mathbf{B}_1^{*}\cdot\tilde{a}\mathbf{E}_0)  dV \nonumber \\
&&-j\omega_ag_{a A B}\epsilon_0 \int_V\left(c^2\mathbf{B}_1^{*}\cdot\tilde{a}\mathbf{B}_{0}-\mathbf{E}_1^{*}\cdot\tilde{a}\mathbf{E}_{0}\right)dV \nonumber \\
&&-\int_V(\mathbf{E}_1^{*}\cdot\mathbf{J}_{e1}^\prime+\mathbf{E}_1^{\prime}\cdot\mathbf{J}_{e1}^*)dV \, ,
\label{PertExp}
\end{eqnarray}
where, $\delta\omega_1=\omega^{\prime}_1-\omega_1$. Now to apply perturbation theory we set all perturbed fields and currents to approximately their unperturbed values, and given that $\frac{1}{\mu_0}\int_V\nabla\cdot\left(\mathbf{E}_1^{*}\times \mathbf{B}_1+\mathbf{E}_1\times\mathbf{B}_1^{*}\right)dV~=-\int_V(\mathbf{E}_1^{*}\cdot\mathbf{J}_{e1}+\mathbf{E}_1\cdot\mathbf{J}_{e1}^*) dV$ and $4U_1=\int_V (\epsilon_0\mathbf{E}_1^{*}\cdot\mathbf{E}_1+\frac{1}{\mu_0}\mathbf{B}_1^{*}\cdot\mathbf{B}_1)dV$, where $U_1$ is the stored energy in the resonator, then from (\ref{PertExp}) we derive
\begin{equation}
\begin{aligned}
\frac{\delta\omega_1}{\omega_1}&\approx\frac{\omega_a\epsilon_0}{4\omega_1U_1}\Big(\int_V(g_{a\gamma\gamma}c\mathbf{E}_1^{*}\cdot\tilde{a}\mathbf{B}_0-g_{aBB}c\mathbf{B}_1^{*}\cdot\tilde{a}\mathbf{E}_0 \\
&+g_{a A B}\left(c^2\mathbf{B}_1^{*}\cdot\tilde{a}\mathbf{B}_{0}-\mathbf{E}_1^{*}\cdot\tilde{a}\mathbf{E}_{0}\right)dV \Big)
\end{aligned}
\label{PertExp2}
\end{equation}
This equation may be used to calculate resonant cavity frequency shifts based on upconversion.

\subsubsection{Frequency Technique}

In this section we use perturbation analysis via equation (\ref{PertExp2}) to derive the sensitivity of the frequency technique, where the background field mixes with the axion to perturb the frequency of the readout mode. Since the integrals must be real, then in terms of unit vectors, we obtain
\begin{equation}
\begin{aligned}
&\frac{\delta\omega_1}{\omega_1}\approx\frac{\omega_a\epsilon_0E_{00}a_0}{2E_{01}\omega_1}\frac{\frac{1}{V}\int_V(g_{a\gamma\gamma}c\mathbf{e}_1\cdot\mathbf{b}_0-g_{aBB}c\mathbf{b}_1\cdot\mathbf{e}_0)~dV}{\frac{1}{V}\int_V\mathbf{e}_1\cdot\mathbf{e}_1^*~dV}.
\end{aligned}
\label{PertExp3}
\end{equation}
Now given that $E_{01}=\sqrt{\frac{2P_{c1}}{\omega_1\epsilon_0V}}$ and $P_{c1}=\frac{4\beta_1Q_{L1}}{(\beta_1+1)^2}P_{1inc}$,  and by considering the sidebands transferred to the coupling circuit, then taking the root mean square average of both sides, then (\ref{PertExp3}) may be written as,
\begin{equation}
\begin{aligned}
\langle\frac{\delta\omega_1}{\omega_1}\rangle
&=\mathcal{K}\omega_{a\gamma\gamma}~g_{a \gamma\gamma}\left\langle a_{0}\right\rangle+\mathcal{K}\omega_{aBB}~g_{a BB}\left\langle a_{0}\right\rangle
\end{aligned}
\label{PertExp4}
\end{equation}
where 
\begin{equation}
\begin{aligned}
&\mathcal{K}\omega_{a\gamma\gamma}=\frac{1}{2}\frac{\omega_a}{\sqrt{\omega_1\omega_0}}\frac{\sqrt{\beta_{0}(\beta_1+1)Q_{L0}}}{\sqrt{\beta_{1}} \sqrt{Q_{L1}}(\beta_0+1)}\sqrt{\frac{P_{0inc}}{P_{1inc}}}\xi_{10} \\
&\mathcal{K}\omega_{aBB}=-\frac{1}{2}\frac{\omega_a}{\sqrt{\omega_1\omega_0}}\frac{\sqrt{\beta_{0}(\beta_1+1)Q_{L0}}}{\sqrt{\beta_{1}} \sqrt{Q_{L1}}(\beta_0+1)}\sqrt{\frac{P_{0inc}}{P_{1inc}}}\xi_{01},
\end{aligned}
\label{PertExp5}
\end{equation}
and depend on the same normalised overlap functions as in the power technique.

To calculate the signal to noise ratio (SNR) for virialised axion dark matter from the galactic halo, for $i= a\gamma\gamma$ or $i= aBB$, then limits on the axion couplings can be undertaken independently by calculating,
\begin{equation}
SNR\omega_{i}=g_i \frac{|\mathcal{K}\omega_{i}|}{\omega_{a}\sqrt{S_{y1}}}\sqrt{\rho_{a} c^{3}}\left(\frac{t}{\Delta f_a}\right)^{\frac{1}{4}},
\label{SNR}
\end{equation}
where, $S_{y1}$ (1/Hz) is the fractional frequency fluctuations competing with the axion signal in the readout oscillator.

The lowest noise oscillators are frequency stabilised by a phase detection scheme, which in principle is limited by the effective readout system noise temperature $T_{RS}$ of the internal phase detector (and includes the resonator and amplifier noise temperature), which is close to ambient temperature for a well-designed system \cite{Ivanov1998}, and in such a case the oscillator noise will be,
\begin{equation}
\sqrt{S_{y1}}=\frac{\sqrt{k_{b} T_{R S}}}{\sqrt{2} Q_{L1} \sqrt{P_{1inc}}} \frac{(1+\beta_1)}{2 \beta_1}\sqrt{\left(2 Q_{L1} \frac{\delta\omega_a}{\omega_{1}}\right)^{2}+1},
\label{sigma2}
\end{equation}
where $P_{1inc}$ is power incident on the input port to the readout mode. In this configuration, usual operating conditions will require $\beta_1\sim1$,  $\beta_0\sim1$ and $\delta\omega_a\sim0$ then the signal to noise ratios become,
\begin{equation}
\begin{aligned}
SNR\omega_{a\gamma\gamma}\sim g_{a\gamma\gamma}
|\xi_{10}|\sqrt{\frac{2Q_{L0}Q_{L1}P_{0inc}\rho_a c^3}{\omega_1\omega_0k_BT_{RS}}}
\left(\frac{10^6t}{ f_{a}}\right)^{\frac{1}{4}},
\end{aligned}
\label{SNRa}
\end{equation}
and
\begin{equation}
\begin{aligned}
SNR\omega_{aBB}\sim g_{aBB}
|\xi_{01}|\sqrt{\frac{2Q_{L0}Q_{L1}P_{0inc}\rho_a c^3}{\omega_1\omega_0k_BT_{RS}}}
\left(\frac{10^6t}{ f_{a}}\right)^{\frac{1}{4}}.
\end{aligned}
\label{SNRa}
\end{equation}
Note, the power and frequency techniques derive the same signal to noise ratio, which depends on the system noise temperature. Thus, the inherent sensitivity of both techniques is the same, which one is better will depend on which can be configured better experimentally to be less influenced by the relevant noise sources and external systematics.
\endgroup

\subsubsection{Sensitivity to ultra-light axions}

Another upconversion technique worth mentioning is the use of the anyon cavity resonator, which uniquely allows the detection of ultra-light axion dark matter due to the non-zero normalised helicity of the cavity mode, given by, 
\begin{equation}
\begin{aligned}
\mathscr{H}_1=\frac{2 \operatorname{Im}[\int\mathbf{B}_1(\vec{r})\cdot\mathbf{E}^*_1(\vec{r})~dV]}{\sqrt{\int\mathbf{E}_1(\vec{r})\cdot\mathbf{E}_1^*(\vec{r})~dV\int\mathbf{B}_1(\vec{r})\cdot\mathbf{B}_1^*(\vec{r})~dV}},
\end{aligned}
\label{helicity}
\end{equation}
so the single mode may act as its own background field \cite{Anyon22,sym14102165}. This technique has been detailed in \cite{Anyon22} and included the QEMD terms to show that this technique was sensitive to the sum of $g_{a\gamma\gamma}$ and $g_{aBB}$, where the helicity is equivalent to the overlap functions defined for the two-mode upconversion detectors. In this case both $g_{a\gamma\gamma}$ and $g_{aBB}$ have the same overlap function.

\section{Discussion}

In this paper we have applied Poynting theorem to axion modified electrodynamics. In QEMD there are three parameters to put limits on $(g_{a\gamma\gamma},g_{aAB},g_{aBB})$, which in principle maybe related if the axion is a QCD axion \cite{SokolovMonopole22} and is a generalisation of the two-photon chiral anomaly if magnetic charge exists. We have shown if the background electromagnetic field is a DC magnetic field then we can only simply configure the experiment to be directly sensitive to $g_{a\gamma\gamma}$. Nevertheless, with a more complicated DC background magnetic field, sensitivity to both $g_{aAB}$ and $g_{a\gamma\gamma}$ may be obtained. Conversely, if the background electromagnetic field is a DC electric field then we can only simply configure the experiment to be directly sensitive to $g_{aAB}$. Nevertheless, with a more complicated DC background electric field, sensitivity to both $g_{aAB}$ and $g_{aBB}$ may be obtained. However, if the background field is an oscillating electromagnetic field, the experiment can be sensitive to $g_{a\gamma\gamma}$ and $g_{aBB}$ at the same time, as we have shown for the upconversion experiments. 

Note, the upconversion experiments search mass ranges less than a $\mu eV$, which is a much lower mass range when compared to the DC Sikivie haloscope \cite{Cat21}. In particular, the anyon cavity haloscope is uniquely sensitive to ultra-light axions \cite{Anyon22}. Other ways to search for axions in the low mass range include reactive axion haloscopes, such as those based on capacitors \cite{tobar2021abraham,Samsonov2022} or inductors \cite{Tong22}, however they were not considered in this work.

The caveat is that we have treated all the axion photon couplings to the axion field, $a$, as independent, and then for each of the couplings the terms $\propto \vec{\nabla} a$ may be omitted. However, it was shown in \cite{SokolovMonopole22} that $\vec{\nabla}{a}$ multiplied by  $g_{aBB}$ can be of the same order of magnitude as $\partial_t{a}$ multiplied by $g_{aAB}$, which could lead to extra sensitivity than to what is calculated in this paper.

\noindent\textbf{Acknowledgments}

This work was funded by the Australian Research Council Centre of Excellence for Engineered Quantum Systems, CE170100009 and Centre of Excellence for Dark Matter Particle Physics, CE200100008. MET acknowledges the Mainz Institute for Theoretical Physics (MITP) of the Cluster of Excellence PRISMA (Project ID 39083149), for enabling a significant portion of this work to be completed. AVS is funded by the UK Research and Innovation grant MR/V024566/1. AR acknowledges support by the Deutsche Forschungsgemeinschaft (DFG, German Research Foundation) under Germany’s Excellence Strategy – EXC 2121 Quantum Universe–390833306. This work has been partially funded by the Deutsche Forschungsgemeinschaft (DFG, German Research Foundation) - 491245950.

\providecommand{\noopsort}[1]{}\providecommand{\singleletter}[1]{#1}%


\begin{thebibliography}{10}

\bibitem{SokolovMonopole22}
Anton~V. Sokolov and Andreas Ringwald.
\newblock Electromagnetic couplings of axions.
\newblock {\em arXiv:2205.02605 [hep-ph]}, 2022.

\bibitem{PQ1977}
R.~D. Peccei and Helen~R. Quinn.
\newblock Cp conservation in the presence of pseudoparticles.
\newblock {\em Phys. Rev. Lett.}, 38:1440--1443, Jun 1977.

\bibitem{Wilczek1978}
F.~Wilczek.
\newblock Problem of strong $p$ and $t$ invariance in the presence of
  instantons.
\newblock {\em Phys. Rev. Lett.}, 40:279--282, Jan 1978.

\bibitem{Weinberg1978}
Steven Weinberg.
\newblock A new light boson?
\newblock {\em Phys. Rev. Lett.}, 40:223--226, Jan 1978.

\bibitem{wisps}
Joerg Jaeckel and Andreas Ringwald.
\newblock The low-energy frontier of particle physics.
\newblock {\em Annual Review of Nuclear and Particle Science}, 60(1):405--437,
  2010.

\bibitem{K79}
Jihn~E. Kim.
\newblock Weak-interaction singlet and strong $\mathrm{CP}$ invariance.
\newblock {\em Phys. Rev. Lett.}, 43:103--107, Jul 1979.

\bibitem{Kim2010}
Jihn~E. Kim and Gianpaolo Carosi.
\newblock Axions and the strong $cp$ problem.
\newblock {\em Rev. Mod. Phys.}, 82:557--601, Mar 2010.

\bibitem{Zhitnitsky:1980tq}
A.~R. Zhitnitsky.
\newblock On possible suppression of the axion hadron interactions. (in
  russian).
\newblock {\em Sov. J. Nucl. Phys.}, 31:260, 1980.

\bibitem{DFS81}
Michael Dine, Willy Fischler, and Mark Srednicki.
\newblock A simple solution to the strong \{CP\} problem with a harmless axion.
\newblock {\em Physics Letters B}, 104(3):199 -- 202, 1981.

\bibitem{SVZ80}
M.A. Shifman, A.I. Vainshtein, and V.I. Zakharov.
\newblock Can confinement ensure natural \{CP\} invariance of strong
  interactions?
\newblock {\em Nuclear Physics B}, 166(3):493 -- 506, 1980.

\bibitem{Dine1983}
Michael Dine and Willy Fischler.
\newblock The not-so-harmless axion.
\newblock {\em Physics Letters B}, 120(1):137 -- 141, 1983.

\bibitem{Preskill1983}
John Preskill, Mark~B. Wise, and Frank Wilczek.
\newblock Cosmology of the invisible axion.
\newblock {\em Physics Letters B}, 120(1):127 -- 132, 1983.

\bibitem{Sikivie1983}
L.F. Abbott and P.~Sikivie.
\newblock A cosmological bound on the invisible axion.
\newblock {\em Physics Letters B}, 120(1--3):133 -- 136, 1983.

\bibitem{Sikivie1983b}
J.~Ipser and P.~Sikivie.
\newblock Can galactic halos be made of axions?
\newblock {\em Phys. Rev. Lett.}, 50:925--927, Mar 1983.

\bibitem{Co2020}
Raymond~T. Co, Lawrence~J. Hall, and Keisuke Harigaya.
\newblock Axion kinetic misalignment mechanism.
\newblock {\em Phys. Rev. Lett.}, 124:251802, Jun 2020.

\bibitem{Co2020b}
Raymond~T. Co and Keisuke Harigaya.
\newblock Axiogenesis.
\newblock {\em Phys. Rev. Lett.}, 124:111602, Mar 2020.

\bibitem{Co2021}
Raymond~T. Co, Lawrence~J. Hall, and Keisuke Harigaya.
\newblock Predictions for axion couplings from alp cogenesis.
\newblock {\em Journal of High Energy Physics}, 2021(1):172, 2021.

\bibitem{Sikivie2021}
Pierre Sikivie.
\newblock Invisible axion search methods.
\newblock {\em Rev. Mod. Phys.}, 93:015004, Feb 2021.

\bibitem{tobar2021abraham}
Michael~E. Tobar, Ben~T. McAllister, and Maxim Goryachev.
\newblock Poynting vector controversy in axion modified electrodynamics.
\newblock {\em Phys. Rev. D}, 105:045009, Feb 2022.
\newblock [Erratum: Phys. Rev. D 106, 109903(E) (2022)].

\bibitem{Cabibbo1962}
N.~Cabibbo and E.~Ferrari.
\newblock Quantum electrodynamics with dirac monopoles.
\newblock {\em Il Nuovo Cimento (1955-1965)}, 23(6):1147--1154, Mar 1962.

\bibitem{Zwanziger1971}
Daniel Zwanziger.
\newblock Local-lagrangian quantum field theory of electric and magnetic
  charges.
\newblock {\em Phys. Rev. D}, 3:880--891, Feb 1971.

\bibitem{TobarModAx19}
Michael~E. Tobar, Ben~T. McAllister, and Maxim Goryachev.
\newblock Modified axion electrodynamics as impressed electromagnetic sources
  through oscillating background polarization and magnetization.
\newblock {\em Physics of the Dark Universe}, 26:100339, 2019.

\bibitem{TOBAR2020}
Michael~E. Tobar, Ben~T. McAllister, and Maxim Goryachev.
\newblock Broadband electrical action sensing techniques with conducting wires
  for low-mass dark matter axion detection.
\newblock {\em Physics of the Dark Universe}, 30:100624, 2020.

\bibitem{sokolov2022gravitational}
Anton~V. Sokolov.
\newblock Gravitational wave electrodynamics.
\newblock {\em arXiv:2203.03278}, 2022.

\bibitem{domcke2022novel}
Valerie Domcke, Camilo Garcia-Cely, and Nicholas~L. Rodd.
\newblock Novel search for high-frequency gravitational waves with low-mass
  axion haloscopes.
\newblock {\em Phys. Rev. Lett.}, 129:041101, Jul 2022.

\bibitem{Singleton:1995dp}
D.~Singleton.
\newblock Topological electric charge.
\newblock {\em International Journal of Theoretical Physics},
  34(12):2453--2466, 1995.

\bibitem{Singleton96}
D.~Singleton.
\newblock Electromagnetism with magnetic charge and two photons.
\newblock {\em American Journal of Physics}, 64(4):452--458, 1996.

\bibitem{Keller2018}
Ole Keller.
\newblock Electrodynamics with magnetic monopoles: Photon wave mechanical
  theory.
\newblock {\em Phys. Rev. A}, 98:052112, Nov 2018.

\bibitem{Rajantie2012}
Arttu Rajantie.
\newblock Introduction to magnetic monopoles.
\newblock {\em Contemporary Physics}, 53(3):195--211, 2012.

\bibitem{Mignaco2001}
Juan~A. Mignaco.
\newblock Electromagnetic duality, charges, monopoles, topology,.
\newblock {\em Brazilian Journal of Physics}, 31(2):235--246, 2001.

\bibitem{Tobar2022b}
Michael~E. Tobar, Raymond~Y. Chiao, and Maxim Goryachev.
\newblock Active electric dipole energy sources: Transduction via electric
  scalar and vector potentials.
\newblock {\em Sensors}, 22(18), 2022.

\bibitem{RHbook2012}
Roger~E. Harrington.
\newblock {\em Introduction to Electromagnetic Engineering}.
\newblock Dover Publications, Inc., 31 East 2nd Street, Mineola, NY 11501, 2nd
  edition, 2012.

\bibitem{Balanis2012}
Constantine~A Balanis.
\newblock {\em Advanced Engineering Electromagnetics}.
\newblock John Wiley, 2012.

\bibitem{Kinsler_2009}
Paul Kinsler, Alberto Favaro, and Martin~W McCall.
\newblock Four poynting theorems.
\newblock {\em European Journal of Physics}, 30(5):983--993, jul 2009.

\bibitem{hagmann1990}
C.~Hagmann, P.~Sikivie, N.~Sullivan, D.~B. Tanner, and S.-I. Cho.
\newblock Cavity design for a cosmic axion detector.
\newblock {\em Review of Scientific Instruments}, 61(3):1076--1085, 1990.

\bibitem{Hagmann1990b}
C.~Hagmann, P.~Sikivie, N.~S. Sullivan, and D.~B. Tanner.
\newblock Results from a search for cosmic axions.
\newblock {\em Physical Review D}, 42(4):1297--1300, August 1990.

\bibitem{Bradley2003}
R.~Bradley, J.~Clarke, D.~Kinion, L.~J. Rosenberg, K.~van Bibber, S.~Matsuki,
  M.~Muck, and P.~Sikivie.
\newblock {Microwave cavity searches for dark-matter axions}.
\newblock {\em Rev. Mod. Phys.}, 75:777--817, 2003.

\bibitem{ADMXaxions2010}
S.~J. Asztalos, G.~Carosi, C.~Hagmann, D.~Kinion, K.~van Bibber, M.~Hotz, L.~J
  Rosenberg, G.~Rybka, J.~Hoskins, J.~Hwang, P.~Sikivie, D.~B. Tanner,
  R.~Bradley, and J.~Clarke.
\newblock Squid-based microwave cavity search for dark-matter axions.
\newblock {\em Phys. Rev. Lett.}, 104:041301, Jan 2010.

\bibitem{ADMX2011}
J.~Hoskins, J.~Hwang, C.~Martin, P.~Sikivie, N.~S. Sullivan, D.~B. Tanner,
  M.~Hotz, L.~J Rosenberg, G.~Rybka, A.~Wagner, S.~J. Asztalos, G.~Carosi,
  C.~Hagmann, D.~Kinion, K.~van Bibber, R.~Bradley, and J.~Clarke.
\newblock Search for nonvirialized axionic dark matter.
\newblock {\em Phys. Rev. D}, 84:121302, Dec 2011.

\bibitem{Braine2020}
T.~Braine, R.~Cervantes, N.~Crisosto, N.~Du, S.~Kimes, L.~J. Rosenberg,
  G.~Rybka, J.~Yang, D.~Bowring, A.~S. Chou, R.~Khatiwada, A.~Sonnenschein,
  W.~Wester, G.~Carosi, N.~Woollett, L.~D. Duffy, R.~Bradley, C.~Boutan,
  M.~Jones, B.~H. LaRoque, N.~S. Oblath, M.~S. Taubman, J.~Clarke, A.~Dove,
  A.~Eddins, S.~R. O'Kelley, S.~Nawaz, I.~Siddiqi, N.~Stevenson, A.~Agrawal,
  A.~V. Dixit, J.~R. Gleason, S.~Jois, P.~Sikivie, J.~A. Solomon, N.~S.
  Sullivan, D.~B. Tanner, E.~Lentz, E.~J. Daw, J.~H. Buckley, P.~M. Harrington,
  E.~A. Henriksen, and K.~W. Murch.
\newblock Extended search for the invisible axion with the axion dark matter
  experiment.
\newblock {\em Phys. Rev. Lett.}, 124:101303, Mar 2020.

\bibitem{Bartram2021}
C.~Bartram, T.~Braine, R.~Cervantes, N.~Crisosto, N.~Du, G.~Leum, L.~J.
  Rosenberg, G.~Rybka, J.~Yang, D.~Bowring, A.~S. Chou, R.~Khatiwada,
  A.~Sonnenschein, W.~Wester, G.~Carosi, N.~Woollett, L.~D. Duffy,
  M.~Goryachev, B.~McAllister, M.~E. Tobar, C.~Boutan, M.~Jones, B.~H. LaRoque,
  N.~S. Oblath, M.~S. Taubman, John Clarke, A.~Dove, A.~Eddins, S.~R. O'Kelley,
  S.~Nawaz, I.~Siddiqi, N.~Stevenson, A.~Agrawal, A.~V. Dixit, J.~R. Gleason,
  S.~Jois, P.~Sikivie, J.~A. Solomon, N.~S. Sullivan, D.~B. Tanner, E.~Lentz,
  E.~J. Daw, M.~G. Perry, J.~H. Buckley, P.~M. Harrington, E.~A. Henriksen, and
  K.~W. Murch.
\newblock Axion dark matter experiment: Run 1b analysis details.
\newblock {\em Phys. Rev. D}, 103:032002, Feb 2021.

\bibitem{McAllister2017}
Ben~T McAllister, Graeme Flower, Eugene~N Ivanov, Maxim Goryachev, Jeremy
  Bourhill, and Michael~E Tobar.
\newblock The organ experiment: An axion haloscope above 15 ghz.
\newblock {\em Phys. Dark Universe}, 18:67--72, 2017.

\bibitem{Quiskamp2022}
Aaron Quiskamp, Ben~T. McAllister, Paul Altin, Eugene~N. Ivanov, Maxim
  Goryachev, and Michael~E. Tobar.
\newblock Direct search for dark matter axions excluding alp cogenesis in the
  63- to 67- micro ev range with the organ experiment.
\newblock {\em Science Advances}, 8(27):eabq3765, 2022.

\bibitem{universe7070236}
Claudio Gatti, Paola Gianotti, Carlo Ligi, Mauro Raggi, and Paolo Valente.
\newblock Dark matter searches at lnf.
\newblock {\em Universe}, 7(7):236, 2021.

\bibitem{10.1093/ptep/ptab051}
Y~Kishimoto, Y~Suzuki, I~Ogawa, Y~Mori, and M~Yamashita.
\newblock {Development of a cavity with photonic crystal structure for axion
  searches}.
\newblock {\em Progress of Theoretical and Experimental Physics}, 2021(6), 04
  2021.
\newblock 063H01.

\bibitem{Devlin2021}
Jack~A. Devlin, Matthias~J. Borchert, Stefan Erlewein, Markus Fleck, James~A.
  Harrington, Barbara Latacz, Jan Warncke, Elise Wursten, Matthew~A. Bohman,
  Andreas~H. Mooser, Christian Smorra, Markus Wiesinger, Christian Will, Klaus
  Blaum, Yasuyuki Matsuda, Christian Ospelkaus, Wolfgang Quint, Jochen Walz,
  Yasunori Yamazaki, and Stefan Ulmer.
\newblock Constraints on the coupling between axionlike dark matter and photons
  using an antiproton superconducting tuned detection circuit in a cryogenic
  penning trap.
\newblock {\em Phys. Rev. Lett.}, 126:041301, Jan 2021.

\bibitem{Kwon2021}
Ohjoon Kwon, Doyu Lee, Woohyun Chung, Danho Ahn, HeeSu Byun, Fritz Caspers,
  Hyoungsoon Choi, Jihoon Choi, Yonuk Chung, Hoyong Jeong, Junu Jeong, Jihn~E.
  Kim, Jinsu Kim, \ifmmode \mbox{\c{C}}\else \c{C}\fi{}a\ifmmode
  \breve{g}\else~\u{g}\fi{}lar Kutlu, Jihnhwan Lee, MyeongJae Lee, Soohyung
  Lee, Andrei Matlashov, Seonjeong Oh, Seongtae Park, Sergey Uchaikin, SungWoo
  Youn, and Yannis~K. Semertzidis.
\newblock First results from an axion haloscope at capp around $10.7\text{
  }\text{ }\ensuremath{\mu}\mathrm{eV}$.
\newblock {\em Phys. Rev. Lett.}, 126:191802, May 2021.

\bibitem{Backes:2021wd}
K.~M. Backes, D.~A. Palken, S.~Al Kenany, B.~M. Brubaker, S.~B. Cahn,
  A.~Droster, Gene~C. Hilton, Sumita Ghosh, H.~Jackson, S.~K. Lamoreaux, A.~F.
  Leder, K.~W. Lehnert, S.~M. Lewis, M.~Malnou, R.~H. Maruyama, N.~M. Rapidis,
  M.~Simanovskaia, Sukhman Singh, D.~H. Speller, I.~Urdinaran, Leila~R. Vale,
  E.~C. van Assendelft, K.~van Bibber, and H.~Wang.
\newblock A quantum enhanced search for dark matter axions.
\newblock {\em Nature}, 590(7845):238--242, 2021.

\bibitem{IWAZAKI2021115298}
Aiichi Iwazaki.
\newblock Axion-radiation conversion by super and normal conductors.
\newblock {\em Nuclear Physics B}, 963:115298, 2021.

\bibitem{Chigusa:2021vh}
So~Chigusa, Takeo Moroi, and Kazunori Nakayama.
\newblock Axion/hidden-photon dark matter conversion into condensed matter
  axion.
\newblock {\em Journal of High Energy Physics}, 2021(8):74, 2021.

\bibitem{PhysRevD.103.096001}
Xunyu Liang, Egor Peshkov, Ludovic Van~Waerbeke, and Ariel Zhitnitsky.
\newblock Proposed network to detect axion quark nugget dark matter.
\newblock {\em Phys. Rev. D}, 103:096001, May 2021.

\bibitem{Alvarez-Melcon:2021aa}
A.~{\'A}lvarez~Melc{\'o}n, S.~Arguedas~Cuendis, J.~Baier, K.~Barth,
  H.~Br{\"a}uninger, S.~Calatroni, G.~Cantatore, F.~Caspers, J.~F. Castel,
  S.~A. Cetin, C.~Cogollos, T.~Dafni, M.~Davenport, A.~Dermenev, K.~Desch,
  A.~D{\'\i}az-Morcillo, B.~D{\"o}brich, H.~Fischer, W.~Funk, J.~D. Gallego,
  J.~M. Garc{\'\i}a~Barcel{\'o}, A.~Gardikiotis, J.~G. Garza, B.~Gimeno,
  S.~Gninenko, J.~Golm, M.~D. Hasinoff, D.~H.~H. Hoffmann, I.~G. Irastorza,
  K.~Jakov{\v c}i{\'c}, J.~Kaminski, M.~Karuza, B.~Laki{\'c}, J.~M. Laurent,
  A.~J. Lozano-Guerrero, G.~Luz{\'o}n, C.~Malbrunot, M.~Maroudas,
  J.~Miralda-Escud{\'e}, H.~Mirallas, L.~Miceli, P.~Navarro, A.~Ozbey,
  K.~{\"O}zbozduman, C.~Pe{\~n}a~Garay, M.~J. Pivovaroff, J.~Redondo, J.~Ruz,
  E.~Ruiz~Ch{\'o}liz, S.~Schmidt, M.~Schumann, Y.~K. Semertzidis, S.~K.
  Solanki, L.~Stewart, I.~Tsagris, T.~Vafeiadis, J.~K. Vogel, E.~Widmann,
  W.~Wuensch, and K.~Zioutas.
\newblock First results of the cast-rades haloscope search for axions at 34.67
  micro ev.
\newblock {\em Journal of High Energy Physics}, 2021(10):75, 2021.

\bibitem{Alesini22}
D.~Alesini, D.~Babusci, C.~Braggio, G.~Carugno, N.~Crescini, D.~D'Agostino,
  A.~D'Elia, D.~Di~Gioacchino, R.~Di~Vora, P.~Falferi, U.~Gambardella,
  C.~Gatti, G.~Iannone, C.~Ligi, A.~Lombardi, G.~Maccarrone, A.~Ortolan,
  R.~Pengo, A.~Rettaroli, G.~Ruoso, L.~Taffarello, and S.~Tocci.
\newblock Search for galactic axions with a high-$q$ dielectric cavity.
\newblock {\em Phys. Rev. D}, 106:052007, Sep 2022.

\bibitem{Sikivie1984}
P~Sikivie.
\newblock Experimental tests of the ``invisible'' axion.
\newblock {\em Phys. Rev. Lett.}, 52(8):695, 1984.

\bibitem{McAllisterFormFactor}
Ben~T. McAllister, Stephen~R. Parker, and Michael~E. Tobar.
\newblock {Axion Dark Matter Coupling to Resonant Photons via Magnetic Field}.
\newblock {\em Phys. Rev. Lett.}, 116(16):161804, 2016.
\newblock [Erratum: Phys. Rev. Lett.117,no.15,159901(2016)].

\bibitem{Samsonov2022}
V.~V. Flambaum, B.~T. McAllister, I.~B. Samsonov, and M.~E. Tobar.
\newblock Searching for scalar field dark matter using cavity resonators and
  capacitors.
\newblock {\em Phys. Rev. D}, 106:055037, Sep 2022.

\bibitem{Goryachev2019}
Maxim Goryachev, Ben~T. McAllister, and Michael~E. Tobar.
\newblock Axion detection with precision frequency metrology.
\newblock {\em Physics of the Dark Universe}, 26:100345, Dec 2019.

\bibitem{Thomson:2021wk}
Catriona Thomson, Maxim Goryachev, Ben~T. McAllister, and Michael~E. Tobar.
\newblock Corrigendum to ``axion detection with precision frequency
  metrology''{$[$}phys. dark universe 26 (2019) 100345{$]$}.
\newblock {\em Physics of the Dark Universe}, 32:100787, 2021.

\bibitem{Cat21}
Catriona~A. Thomson, Ben~T. McAllister, Maxim Goryachev, Eugene~N. Ivanov, and
  Michael~E. Tobar.
\newblock Upconversion loop oscillator axion detection experiment: A precision
  frequency interferometric axion dark matter search with a cylindrical
  microwave cavity.
\newblock {\em Phys. Rev. Lett.}, 126:081803, Feb 2021.
\newblock [Erratum: Phys. Rev. Lett.127, 019901(2021)].

\bibitem{berlin2020axion}
Asher Berlin, Raffaele~Tito D'Agnolo, Sebastian~AR Ellis, Christopher Nantista,
  Jeffrey Neilson, Philip Schuster, Sami Tantawi, Natalia Toro, and Kevin Zhou.
\newblock Axion dark matter detection by superconducting resonant frequency
  conversion.
\newblock {\em Journal of High Energy Physics}, 2020(7):1--42, 2020.

\bibitem{Lasenby2020b}
Robert Lasenby.
\newblock Parametrics of electromagnetic searches for axion dark matter.
\newblock {\em Phys. Rev. D}, 103:075007, Apr 2021.

\bibitem{ABerlin2021}
Asher Berlin, Raffaele~Tito D'Agnolo, Sebastian A.~R. Ellis, and Kevin Zhou.
\newblock Heterodyne broadband detection of axion dark matter.
\newblock {\em Phys. Rev. D}, 104:L111701, Dec 2021.

\bibitem{Lasenby2020}
Robert Lasenby.
\newblock Microwave cavity searches for low-frequency axion dark matter.
\newblock {\em Phys. Rev. D}, 102:015008, Jul 2020.

\bibitem{Cat23}
Catriona~A. Thomson, Maxim Goryachev, Ben~T. McAllister, Eugene~N. Ivanov, Paul
  Altin, and Michael~E. Tobar.
\newblock Searching for low-mass axions using resonant upconversion.
\newblock {\em arXiv:2301.06778 [hep-ex]}, 2023.

\bibitem{HARTNETT2011}
John~G. Hartnett, Joerg Jaeckel, Rhys~G. Povey, and Michael~E. Tobar.
\newblock Resonant regeneration in the sub-quantum regime -- a demonstration of
  fractional quantum interference.
\newblock {\em Physics Letters B}, 698(5):346 -- 352, 2011.

\bibitem{parker2013b}
Stephen~R. Parker, John~G. Hartnett, Rhys~G. Povey, and Michael~E. Tobar.
\newblock Cryogenic resonant microwave cavity searches for hidden sector
  photons.
\newblock {\em Phys. Rev. D}, 88:112004, Dec 2013.

\bibitem{Ivanov1998}
E.~N. {Ivanov}, M.~E. {Tobar}, and R.~A. {Woode}.
\newblock Microwave interferometry: application to precision measurements and
  noise reduction techniques.
\newblock {\em and Frequency Control IEEE Transactions on Ultrasonics,
  Ferroelectrics}, 45(6):1526--1536, November 1998.

\bibitem{Anyon22}
J.~F. Bourhill, E.~C.~I. Paterson, M.~Goryachev, and M.~E. Tobar.
\newblock Twisted anyon cavity resonators with bulk modes of chiral symmetry
  and sensitivity to ultra-light axion dark matter.
\newblock {\em arXiv:2208.01640 [hep-ph]}, 2022.

\bibitem{sym14102165}
Michael~E. Tobar, Catriona~A. Thomson, William~M. Campbell, Aaron Quiskamp,
  Jeremy~F. Bourhill, Benjamin~T. McAllister, Eugene~N. Ivanov, and Maxim
  Goryachev.
\newblock Comparing instrument spectral sensitivity of dissimilar
  electromagnetic haloscopes to axion dark matter and high frequency
  gravitational waves.
\newblock {\em Symmetry}, 14(10), 2022.

\bibitem{Tong22}
Tong Li, Rui-Jia Zhang, and Chang-Jie Dai.
\newblock Solutions to axion electromagnetodynamics and new search strategies
  of sub micro ev axion.
\newblock {\em arXiv:2211.06847 [hep-ph]}, 2022.

\end{thebibliography}
\end{document}